\documentstyle[sprocl,epsfig]{article}

\def\ra{\rightarrow}

\def\beq{\begin{equation}}
\def\eeq{\end{equation}}
\def\bea{\begin{eqnarray}}
\def\eea{\end{eqnarray}}
\newcommand{\lsim}{\raisebox{-0.13cm}{~\shortstack{$<$ \\[-0.07cm] $\sim$}}~}
\newcommand{\gsim}{\raisebox{-0.13cm}{~\shortstack{$>$ \\[-0.07cm] $\sim$}}~}
\newcommand{\ee}{e^+e^-}

\newcommand{\tb}{\tan\beta}

\def\t1{\tilde{t_1}}
\begin{document}
\thispagestyle{empty}
\begin{flushright}
PM/99--40\\
\end{flushright}

\vspace{1cm}

\begin{center}

{\large\sc {\bf HIGGS PARTICLES AT LINEAR e$^+$e$^-$ COLLIDERS}}: 

\vspace*{0.5cm}

{\large\sc {\bf THEORETICAL ISSUES}}

\vspace{1cm}

{\sc Abdelhak Djouadi} 

\vspace{0.5cm}

Laboratoire de Physique Math\'ematique et Th\'eorique, UMR5825--CNRS,\\
Universit\'e de Montpellier II, F--34095 Montpellier Cedex 5, France.

\vspace*{2cm}

{\large\sc {\bf Abstract}}

\end{center}

\vspace*{0.5cm}

\noindent I summarize the prospects for discovering and studying 
the properties of Higgs particles at future high--energy and high--luminosity 
$\ee$ linear colliders. I will focus on the Higgs particle of the Standard 
Model and the Higgs bosons predicted by Supersymmetric theories.

\vspace*{2cm}

\centerline{Plenary talk given at the International Workshops on Linear 
Colliders (LCWS99)}
\centerline{Sitges, Barcelona, Spain, on April 28-May 5, 1999}

\setcounter{page}{0}

\newpage

\title{HIGGS PARTICLES AT LINEAR e$^+$e$^-$ COLLIDERS: \\
THEORETICAL ISSUES}

\author{Abdelhak DJOUADI}

\address{Physique Math\'ematique et Th\'eorique, UMR 5825--CNRS, \\
Universit\'e Montpellier II, F--34095 Montpellier Cedex 5, France}


\maketitle\abstracts{I summarize the prospects for discovering and studying 
the properties of Higgs particles at future high--energy and high--luminosity 
$\ee$ linear colliders. I will focus on the Higgs particle of the Standard 
Model and the Higgs bosons predicted by Supersymmetric theories.} 
\section{Introduction} 
One of the most important missions of future high--energy colliders will be the
search for scalar Higgs particles and the exploration of the electroweak
symmetry breaking mechanism. In the Standard Model (SM), one doublet of
complex scalar fields is needed to spontaneously break the SU(2)$\times$U(1)
symmetry. Among the four initial degrees of freedom, there Goldstones will
be absorbed by the $W^\pm$ and $Z$ bosons to get their masses, and the
remaining degree of freedom will correspond to a physical scalar particle,
the Higgs boson.\cite{x0}

Since the couplings of the Higgs boson to fermions and gauge bosons are
proportional to the masses of these particles, the only unknown parameter in
the SM is the Higgs boson mass, $M_H$. It is a free parameter and the only
things we know about it are that: $i)$ it should be larger than\cite{x1} $\sim
100$ GeV from the negative searches at LEP and $ii)$ it is probably smaller
than $\sim 1$ TeV, since for higher values the electroweak gauge bosons would
interact strongly to insure unitarity in their scattering and perturbation
theory would be lost.\cite{x0}

However, there are both theoretical and experimental hints which indicate that 
the Higgs boson of the SM might be rather light: 

-- Global fits of the electroweak precision observables at LEP, SLC and the
Tevatron favor a Higgs boson [whose loop contributions to the electroweak
parameters depend logarithmically on $M_H$] with a mass around 100 GeV; an
upper bound\cite{x2} of $M_H \lsim 260$ GeV has been set at the $95\%$
confidence level.  

-- The quartic Higgs coupling is proportional to $M_H^2$ and since the scalar
sector of the SM is not an asymptotically free theory, the coupling will grow
with the energy until it reaches the Landau pole, where the theory does not
make sense anymore. If the cut--off $\Lambda$ where  new phenomena should occur
is of ${\cal O}(1$ TeV), the Higgs mass should be smaller than $\sim 500$ GeV
[as verified by simulations on the lattice]. But if one wants to extend the SM
up the GUT scale $\Lambda_{\rm GUT} \sim 10^{16}$ GeV [a prerequisite for the
perturbative renormalization of $\sin^2\theta_W$ from the GUT value 3/8 down to
the experimentally observed value], $M_H$ is restricted to much smaller values.
In addition, radiative corrections due to top quark loops could drive the Higgs
self--coupling to negative values, therefore destabilizing the vacuum.  The
stability and the triviality bounds, constrain the SM Higgs boson mass to lie
in the range:\cite{x3} 130 GeV $\lsim M_H \lsim$ 180 GeV.

However, there are two problems that one has to face, when trying to extend the
SM to $\Lambda_{\rm GUT}$. The first one is the so--called hierarchy or
naturalness problem: the Higgs boson tends to acquire a mass of the order of
the large scale [the radiative corrections to $M_H$ are quadratically
divergent].  The second problem is that the simplest GUTs predict a value for
$\sin^2\theta_W$ that is incompatible with the measured one $\sim 0.23$. Low
energy Supersymmetry (SUSY)\cite{susy} solves these two problems at once: SUSY
particles loops cancel the quadratic divergences to the Higgs boson mass and
contribute to the running of the gauge coupling constants to correct for the
small discrepancy to the observed value of $\sin^2\theta_W$.  

The minimal supersymmetric extension of the Standard Model (MSSM) requires the 
existence of two isodoublets of Higgs fields,\cite{x0} leading to three
neutral, $h/H$ (CP--even with $h$ being the lightest particle), $A$ (CP--odd)
and a pair of charged scalar particles $H^\pm$.  Besides the four masses, two
additional parameters define the properties of these particles: a mixing angle
$\alpha$ in the neutral CP--even sector and $\tb$ the ratio of the two vacuum
expectation values of the Higgs fields, which from GUT restrictions is assumed
in the range $1 < \tb <m_t/m_b$.  Supersymmetry leads to several relations
among these parameters and only two of them [taken in general to be $\tb$ and
$M_A$] are in fact independent.  These relations impose a strong hierarchical
structure of the mass spectrum and lead to the definite prediction that, at the
tree level, the neutral $h$ boson should be lighter that the $Z$ boson.

However, radiative corrections involving mainly the top quark and its SUSY
partners, introduce new [soft SUSY--breaking] parameters in the Higgs sector
and affect the Higgs boson masses and couplings significantly.  The leading
part of these corrections grows as the fourth power of the top quark mass and
logarithmically with the common squark mass [a strong dependence on the
trilinear stop coupling $A_t$ is also present], and shift the mass of the
lightest $h$ boson upwards. A recent calculation,\cite{x4} performed at the
two--loop level in the diagrammatic approach, restrict the $h$ boson mass to be
less than $\sim 135$ GeV.  

Note that for large values of $M_A$, the heavy neutral and charged Higgs bosons
are nearly mass degenerate, while the lightest $h$ boson reaches its maximal
mass value. One is then in the so--called decoupling regime where the lightest
$h$ boson has almost the same properties as the standard Higgs boson [but with
a mass below $\sim 135$ GeV] and the SM and MSSM Higgs sectors look practically
the same.

In more general SUSY scenarii, one can add an arbitrary number of Higgs 
doublet and/or singlet fields without being in conflict with high precision 
data.\cite{x0} The Higgs spectrum becomes then much more complicated than 
in the MSSM, and much less constrained. However, the triviality argument
always imposes a bound on the mass of the lightest Higgs boson of the theory. 
For instance, if only one Higgs singlet field is added to the MSSM, an upper
bound $M_h \lsim 150$ GeV can be derived.\cite{x5} In the most general
SUSY model, with arbitrary matter content and gauge coupling unification 
near the GUT scale, and absolute upper limit on the mass of the lightest 
Higgs boson, $M_h \lsim 205$ GeV, has been recently derived.\cite{x6}

Thus, either in the SM or in its SUSY extensions, a Higgs boson should be
lighter than $\sim 200$ GeV, and will be therefore kinematically accessible
at an $\ee$ linear collider with a c.m. energy $\sqrt{s} \gsim 350$ GeV. In 
this talk, I will summarize the prospects for such a collider to discover and 
to study the properties of this particle.\cite{ee}

\section{Higgs Boson Decays and Production} 

In the SM, the profile of the Higgs particle [decay widths, branching ratios
and production cross sections] is uniquely determined once $M_H$ is fixed.  The
profile of the MSSM Higgs bosons is determined to a large extent also by their
couplings to fermions and gauge bosons, which in general depend strongly on
$\tb$ [and $\alpha$].

\vspace*{-3mm}

\subsection{Higgs decays} 
In the ``low mass" range $M_H\lsim 130$ GeV, the SM Higgs boson 
decays\cite{decay}
into a large variety of channels, the main mode being by far into $b\bar{b}$ 
pairs with a BR of $\sim 90\%$ followed by the decays into $c\bar{c}$ and 
$\tau^+\tau^-$ pairs with BRs of $\sim 5\%$. Also of significance is  the 
top--loop mediated Higgs decay into gluons which for $M_H \sim$ 120 GeV occurs 
at the level of $\sim 5\%$. The top and $W$--loop mediated $\gamma\gamma$ and 
$Z \gamma$ decay modes are very rare, the BRs being of ${\cal O }(10^{-3})$; 
however the $\gamma \gamma$ decays lead to clear signals and are interesting 
being sensitive to new heavy particles.  Note that QCD
corrections to the hadronic decays turn out be quite substantial, and together
with the rather imprecise present knowledge of the strong coupling constant
$\alpha_s$ and the $c$ and $b$ quark masses, introduce some uncertainties in 
the BRs.

In the ``high mass" range $M_H \gsim 130$ GeV, the $H$ bosons decay into 
$WW$ and $ZZ$ pairs, with one of the gauge bosons being virtual below the 
threshold. Above the $ZZ$ threshold, the Higgs boson decays almost 
exclusively into these channels with BRs of 2/3 for $WW$ and 1/3 for $ZZ$ 
[for high $M_H$ values, the opening of the $t\bar{t}$ channel does not alter
significantly this pattern].

In the low mass range, the $H$ boson is very narrow $\Gamma_H \lsim 10$ MeV,
but the width becomes rapidly wider for masses larger than 130 GeV,
reaching 1 GeV at the $ZZ$ threshold. The Higgs decay width cannot be
measured directly for $M_H \lsim 200$ GeV. 

The decay pattern of the MSSM Higgs bosons depends strongly on $\tb$. 
For large $\tb$ values, it is simple a result of the strong enhancement of the
Higgs couplings to down--type fermions: the neutral Higgs bosons will decay
into $b\bar{b}$ ($\sim 90\%$) and $\tau^+ \tau^-$ ($\sim 10\%)$ pairs, and
$H^\pm$ into $\tau \nu_\tau$ pairs below and $tb$ pairs above the top--bottom
threshold.  Only when $M_h$ approaches its maximal value is this simple rule
modified since in this decoupling limit, the $h$ boson decays as the SM Higgs
boson. For small values of $\tb$, the decay pattern of the heavy neutral
Higgs bosons can be more complicated. The $b$ decays are in general
not dominant any more; instead, cascade decays to pairs of light Higgs
bosons and mixed pairs of Higgs and gauge bosons are important and
decays to $WW/ZZ$ pairs will play a role. For very large masses, they decay 
almost exclusively to top quark pairs. The decay pattern of the charged 
Higgs bosons for small $\tb$ is similar to that at large $\tb$ except in 
the intermediate mass range where cascade decays to $Wh$ are dominant.
In addition, below threshold three--body decays might be important.

When the decays into SUSY particles are kinematically allowed [for the heavy
Higgs scalars] the pattern becomes even more complicated since the decay 
channels into charginos, neutralinos and squarks might be 
non--negligible. 

In more general SUSY scenarii, the decays of the Higgs bosons can be much more 
complicated than in the MSSM; however, this do not lead to any difficulty to 
detect some of the particles at $\ee$ colliders as will be discussed later. 

\subsection{Higgs production}
The main production mechanism\cite{ee} 
of the SM Higgs particles in $\ee$ collisions are
the Higgs--strahlung process, $\ee \ra (Z^*) \ra ZH$ [with a cross section
which scales as $1/s$ and therefore dominates at low energies], and the $WW$
fusion mechanism, $\ee \ra \nu \bar{\nu} (W^* W^*)\ra \nu \bar{\nu} H$ [with a
cross section rising like $\log(s/M_H^2)$ and which dominates at high
energies]. The cross section for the $ZZ$ fusion mechanism, $\ee \ra \ee H$, is
an order of magnitude smaller than the later due to the smallness of the NC
couplings compared to the CC ones, but gives some complementary 
information.\cite{mink}  There are also higher order processes:
associated Higgs production with top quarks or a photon and
double Higgs production in the strahlung and fusion processes or through
loops; they have smaller cross sections but are very useful when
it comes to study the Higgs properties as will be discussed later. Additional
production mechanisms are also provided by the $\gamma \gamma \ra H$ and $e
\gamma \ra \nu WH$ processes, the high--energy photons generated by Compton
back scattering of laser light; they are discussed elsewhere.\cite{x11}

At $\sqrt{s} \sim 500$ GeV, the Higgs--strahlung and the $WW$ fusion processes 
have approximately the same cross sections for the mass range 100 GeV $\lsim M_H
\lsim$ 200 GeV. With a luminosity $\int {\cal L} \sim 500$ fb$^{-1}$ as it is
expected for the TESLA design,\cite{x10} a sample of $\sim$ 75.000 Higgs bosons
can be collected in a one year running for $M_H\sim 130$ GeV. Assuming that 25
events are required to establish a discovery [the signal with $H \ra b\bar{b}$
is easy to detect at $\ee$ colliders, especially with efficient micro--vertex
detectors\cite{x12}] less than one hour running is needed in such a machine 
[to be compared with the much longer running time at the LHC for the $H \ra 
\gamma \gamma$ mode\cite{pp}]. Thus, the discovery of the SM Higgs particle
[if kinematically allowed] is not a problem in the clean environment of such 
an $\ee$ collider.  

In the MSSM, besides the usual Higgs--strahlung and fusion processes for the
production of the CP--even Higgs bosons $h$ and $H$, the neutral Higgs
particles can also be produced pairwise: $\ee \ra A + h/H$. The cross
sections for the Higgs--strahlung and the pair production as well as the
cross sections for the production of $h$ and $H$ are mutually complementary,
coming either with a coefficient $\sin^2(\beta- \alpha)$ or $\cos^2(\beta
-\alpha)$. The sum of the cross sections for $h$ production in the strahlung
and associated processes is roughly the same as the cross section for the
SM Higgs boson [with the same mass] in the strahlung process.  
The CP--even Higgs particles can also be searched for in the $WW$ and $ZZ$
fusion mechanisms. Charged Higgs bosons can be produced pairwise, $\ee \ra
H^+H^-$, through $\gamma,Z$ exchange and the cross section which depends only 
on $M_{H^\pm}$, is large up to $M_{H^\pm} \sim 230$~GeV; the $H^\pm$ bosons  
can also be produced in top decays if kinematically allowed.  

The discussion on the MSSM Higgs production at $\ee$ linear colliders can be
summarized in the following points\cite{ee}: $i)$ The Higgs boson $h$ can be 
detected in
the entire range of the MSSM parameter space, either through the bremsstrahlung
process or through pair production; in fact, this conclusion holds true even at
a c.m. energy of 300 GeV and with a luminosity of a few fb$^{-1}$. $ii)$ All
SUSY Higgs bosons can be discovered at a 500 GeV collider if the $H,A$ and
$H^{\pm}$ masses are less than $\sim 230$ GeV; for higher masses, one simply
has to increase the c.m.  energy. $iii)$ Even if the decay modes of the 
Higgs bosons are very complicated [e.g. they decay invisibly] missing mass
techniques allow their detection.  

In extensions of the MSSM, the Higgs production processes are as the ones above
but the phenomenological analyses are more involved since there is more 
freedom in the choice of parameters. However, even if the Higgs sector is 
extremely complicated, there is always a light Higgs boson which has sizeable
couplings to the $Z$ boson. This Higgs particle can be thus produced in the
strahlung process, $\ee \ra Z+$``$h$", and using the missing mass technique this
``$h$" particle can be detected. Recently a ``no--loose theorem" has been 
proposed:\cite{no-loose}  a Higgs boson in SUSY theories can be always detected
at a 500 GeV $\ee$ collider with a luminosity of $\int {\cal L} \sim 500$ fb
$^{-1}$ in the strahlung process, regardless of the complexity of the Higgs
sector of the theory and of the decays of the Higgs boson.

\section{Precision Measurements at the LC}

Thus a light Higgs boson can be found without any problem at a future linear
collider. However, such a particle might be first discovered at the present
machines LEP\cite{lep} and the Tevatron,\cite{tev} or at the LHC.\cite{pp} As
discussed by M. Peskin in the introductory talk,\cite{x14} the job of a linear
$\ee$ collider, will be rather, to study the properties of the Higgs particles.
The clean environment and the very high luminosities which are expected [e.g.
$\int {\cal L} \gsim 100$ fb$^{-1}$ for the TESLA design], allow to study these
properties in great details and to make very accurate measurements in the Higgs
sector.  We summarize below the measurements which can be made in the main
production mechanisms as well as in the higher--order processes. We will focus
on the case of the SM Higgs boson, which is equivalent for $M_H \lsim 130$ GeV,
to the case of the light $h$ boson of the MSSM close to the decoupling regime. 
A more quantitative discussion will be given by S. Yamashita.\cite{x12}  

\vspace*{-3mm}

\subsection{Measurements in the main processes}

\hspace*{4mm} $\bullet$ The measurement of the recoil $\ee$ or $\mu^+ \mu^-$
mass in the Higgs--strahlung process, $\ee \ra ZH\ra He^+e^-$ and $H\mu^+
\mu^-$, allows 
a very good determination of the Higgs boson mass. At $\sqrt{s}=350$ GeV and
with a luminosity of $\int {\cal L}= 500$ fb$^{-1}$, a precision of $\sim 150$
MeV can be reached\cite{Lohmann-Garcia-Abia} for a Higgs boson mass of $M_H
\sim 120$ GeV.  The precision can be significantly increased if one uses 
the hadronic decays of the $Z$ boson [which have more statistics] but
since the mass resolution is rather bad in the simplest way, one has to make
some kinematical fits of 4--jets with distributions.\cite{juste} A threshold
scan might also improve the measurement. The one per
mile accuracy which can be obtained for the Higgs boson mass can be very
important, especially in the MSSM where it allow to strongly constrain the
other parameters of the model.

$\bullet$ The angular distribution of the $Z/H$ in the Higgs--strahlung process
is sensitive to the spin--zero of the Higgs particle: at high--energies the $Z$
is longitudinally polarized and the distribution follows the $\sim
\sin^2\theta$ law which unambiguously characterizes the production of a
$J^P=0^+$ particle. The spin--parity quantum numbers of the Higgs bosons can
also be checked experimentally by looking at correlations in the production
$\ee \ra HZ \ra$ 4--fermions or decay $H \ra WW^* \ra$ 4--fermions processes, as
well as in the more difficult channel $H \ra \tau^+ \tau^-$ for $M_H \lsim 140$
GeV. An unambiguous test of the CP nature of the Higgs bosons can be made in
the process $\ee \ra t \bar{t}H$ or at laser photon colliders in the loop
induced process $\gamma \gamma \ra H$.  

$\bullet$ The masses of the gauge bosons are generated through the Higgs
mechanism and the Higgs couplings to these particles are proportional to their
masses.  This fundamental prediction has to be verified experimentally.  The
Higgs couplings to $ZZ/WW$ bosons can be directly determined by measuring the
production cross sections in the bremsstrahlung and the fusion processes.  In
the $\ee \ra H \ee$ and $H\mu^+\mu^-$ processes, the total cross section can be
measured\cite{Lohmann-Garcia-Abia} with a precision less than $\sim 3\%$ at 
$\sqrt{s}=350$ GeV and with $\int {\cal L}= 500$ fb$^{-1}$. This leads to an 
accuracy of $\sim 1.5\%$ on the $HZZ$ coupling.  

$\bullet$ The measurement of the branching ratios of the Higgs boson are of
utmost importance. For Higgs masses below $M_H \lsim 130$ GeV a large variety of
BRs can be measured at the linear collider. The $b\bar{b}, c\bar{c}$  and
$\tau^+ \tau^-$ BRs allow to measure the relative couplings of the Higgs
bosons to these fermions and to check the fundamental prediction of the Higgs
mechanism that they are proportional to fermion masses. In particular BR$(H \ra
\tau^+ \tau^-) \sim m_{\tau}^2/3\bar{m}_b^2$ allows to make such a test.  In
addition, these branching ratios, if measured with enough accuracy, could allow
to distinguish a Higgs boson in the SM from its possible extensions. The gluonic
BR is sensitive to the $t\bar{t}H$ Yukawa coupling [and might therefore give an
indirect measurement of this important coupling] and to new strongly
interacting particles which couple to the Higgs boson [such as top squarks in
SUSY extensions of the SM].  The branching ratio into $W$ boson starts to
be significant for Higgs masses of the order of 120 GeV and allows to measure 
the $HWW$ coupling. The BR of the loop induced $\gamma \gamma$ decay of the 
Higgs boson is also very important since it is sensitive to new particles
[the measurement of this BR gives the same information as the measurement of
the cross section for Higgs boson production at $\gamma \gamma$ colliders]. 

In this workshop, a lot of experimental work has been performed  to assess 
the level of precision with which all these BRs can be measured,\cite{x12} 
and the results are very impressive. Table~1 from Ref.\cite{richard} summarizes
the achieved precision
at $\sqrt{s}=350$ GeV and with $\int {\cal L}= 500$ fb$^{-1}$ [details can be 
found in\cite{x12}]. These errors are so small that one can tell a SM Higgs
 boson from the MSSM $h$ boson [whose couplings to fermions and gauge bosons 
are in principle altered by mixing angle factors] up to a pseudoscalar Higgs 
boson mass of $M_A \sim 700$ GeV. In fact, the experimental errors are even 
smaller than the theoretical errors which affect some BRs [in particular for 
the $gg$ and $c\bar{c}$ modes\cite{richard}] due to the [mostly 
experimental...] uncertainties in the measurement of $\alpha_s, m_c$ and to a 
lesser extent $m_b$.  

\smallskip

\begin{center}
\begin{tabular}{|c|c|c|c|c|c|} \hline
$b\bar{b}$ & $c\bar{c}$ & $gg$ & $\tau^+ \tau^-$ & $W^+ W^-$ & $\gamma \gamma$
\\ \hline  
2\% & 8\% & 6\% & 6\% & 2--10\% & 20\% \\ \hline
\end{tabular}
\end{center}
\vspace{0.3cm}
\noindent {\small Table 1: Expected accuracies on Higgs BR's at $\sqrt{s}
350$ GeV and with $\int {\cal L}= 500$ fb$^{-1}$.}

\smallskip

$\bullet$ As discussed previously, the total width of the Higgs boson [for 
masses less than $\sim 200$ GeV] is so small that it cannot be resolved 
experimentally. However, the measurement of BR($H \ra WW$) allows an indirect
determination of $\Gamma_H$ since the $HWW$ coupling can be determined from
the measurement of the Higgs production cross section in the $WW$ fusion 
process [or from the measurement of the cross section of the Higgs--strahlung 
process, assuming SU(2) invariance]. The accuracy of the $\Gamma_H$ measurement
follows then from that of the $WW$ branching ratio.\cite{richard}

\subsection{Measurements in higher order processes}
There are several processes where Higgs particles are produced in pairs or in 
association with heavy particles or else through loop diagrams. Since these 
processes are of higher order in perturbation theory, the production rates are 
in general rather small, at or below the femtobarn level. Very high luminosities
$\int {\cal L} \sim 1$ ab$^{-1}$ offer a unique opportunity to study these
processes and to gain additional information on the Higgs sector. Some of 
these processes have been discussed in the parallel sessions and the main
points are summarized below: 

$\bullet$ Associated production of Higgs bosons with top quark pairs $\ee \ra
t\bar{t}H$: \\ 
The Higgs coupling to top quarks, which is the largest coupling in the
electroweak SM, is directly accessible in this process.\cite{ttH} In addition
to Higgs radiation from the quark lines which gives access to the $t\bar{t}H$
Yukawa coupling, there is also Higgs emission from the $Z$ line [and diagrams
with the exchange of heavier Higgs bosons in the MSSM], which nevertheless give
small contributions to the production cross section.  The later is at the
femtobarn level for $M_H \sim 100$ GeV at a c.m.  energy of 500 GeV, but the
signal is quite spectacular [two $W$ bosons and four $b$ quarks, with
kinematical constraints to reconstruct the top quarks and the $H$ boson] giving
the possibility of isolating these events with a luminosity of ${\cal O}(1$
ab$^{-1}$). A recent analysis,\cite{ttHexp} with detailed simulations of the
signal and backgrounds including realistic detector effects and reconstruction
procedures, has shown that an accuracy of $\sim 5\%$ can be achieved in the
measurement of the $t \bar{t}H$ coupling for a mass $M_H\simeq 120$ GeV at a
c.m.  energy $\sqrt{s}= 800$ GeV and with a luminosity $\int {\cal L}=1$
ab$^{-1}$. The QCD corrections to this process have been calculated
recently.\cite{ttHth} They are large and positive [with $K$--factors of the
order of 1.4 to 2.4] at $\sqrt{s} \sim 500$ GeV because of resonance effects,
and small and negative [with $K$--factors of order 0.8--0.9] at $\sqrt{s} \sim
1$ TeV. Note that the associated production of Higgs bosons with $b\bar{b}$
pairs can have a significant cross section in the MSSM, for large $\tb$ values
and a low pseudoscalar $A$ mass; in this case this processes would allow a nice
direct determination\cite{sopzack} of the important parameter $\tb$ [which is
difficult to achieve in other processes]. 

$\bullet$ Double Higgs production in the strahlung process $\ee \ra H H Z $: \\
To establish the Higgs mechanism experimentally in an unambiguous way, the
self--energy potential of the Higgs field must be reconstructed. This requires
the determination of the trilinear [and quadrilinear] self--couplings as
predicted for instance in the SM or MSSM. These couplings can be probed in the
production of pairs of neutral Higgs bosons. A coherent picture of the
trilinear couplings has been given here\cite{kilian} with the production of
pairs of neutral Higgs bosons in the SM and MSSM, in all relevant channels of
double Higgs--strahlung, associated multi--Higgs production and $WW/ZZ$ fusion
to Higgs boson pairs. The most interesting process at energies around 500 GeV
is the double Higgs--strahlung process, $\ee \ra HHZ$. The cross section, which
is very sensitive to the trilinear self--coupling, is of the order 0.5 fb for
$M_H \sim 100$ GeV. This leads to approximately one thousand events for a
luminosity of $\int {\cal L}=2$ ab$^{-1}$ [corresponding to 4 years of running
with the expected luminosity at TESLA] with an extremely clean signal [a $Z$
boson with 4 $b$--jets with two $b\bar{b}$ pairs having an invariant mass $M_H$
which is expected to be measured precisely in the main production processes]. A
detailed simulation\cite{lutz} has shown that the trilinear coupling can be
measured with a precision at the $\sim 15\%$ level. This analysis is
preliminary and not yet optimized, and a better determination is to be expected
in the future. At higher energies, double Higgs production in the $WW$ fusion
channel, $\ee \ra \nu \bar{\nu}HH$, which has a larger cross section [$\sim 1$
fb at $\sqrt{s}=1.5$ TeV for $M_H \sim 100$ GeV], might be used. The
quadrilinear Higgs self--coupling can be measured in triple Higgs boson
production, but the cross section is suppressed by an additional electroweak
factor, and is therefore too small to be observable.\cite{triple}

$\bullet$ Associated production of a Higgs boson and a photon, $\ee \ra 
H\gamma$: \\
In the SM, this process proceeds through $s$--channel $\gamma^* \gamma H$ and
$Z^* \gamma H$ vertex diagrams, but additional $t$--channel vertex and box
diagrams involving $W$/neutrino and $Z$/electron exchange also occur. It is
therefore sensitive to the $H \gamma \gamma$ and $H Z \gamma$ vertices. These
couplings do not occur at the tree level but are induced by loops of heavy
particles, which if their interaction with the Higgs boson is proportional to
their masses, do not decouple for very large masses. These vertices could
therefore serve to count the number of particles which couple to the $H$ boson.
[These couplings can be also accessed in the decays $H \ra \gamma \gamma$ and
$H\ra Z \gamma$ but the BRs are very small $\sim 10^{-3}$; the $H\gamma \gamma$
coupling can also be determined directly by means of the laser $\gamma\gamma \ra
H$ fusion process]. A precise determination of these couplings could help to
distinguish between the SM Higgs boson and Higgs particles predicted by some of
its extensions such a two--Higgs Doublet Model\cite{maria} [in these models, it
is also useful for $h$ and $A$ production in the process $Z \ra \gamma$+ Higgs
with the linear collider running on the $Z$--resonance,\cite{maria} since the
experimental bounds on the masses are not as tight as in the MSSM] or
supersymmetric theories\cite{nous} [where scalar tops and charginos loops might
have some significant contributions]. Unfortunately, the cross section is
rather small: at a 500 GeV collider it is of ${\cal O}(0.1$ fb), leading to two
hundred events events with $\int {\cal L}=2$ ab$^{-1}$. This number would 
allow, roughly, a measurement of the cross section at the $10\%$ level. The
monochromatic photon makes the signal very clean, but not detailed simulation
has been performed yet to access the viability of this signal.

$\bullet$ Associated production of Higgs bosons with top squark pairs, $\ee \ra
\tilde{t} \tilde{t}h$: \\ 
In the MSSM, if the mixing between third generation squarks is large, scalar
top [and also bottom] quarks can be rather light and at the same time, their
coupling to Higgs bosons can become substantial. For instance, the couplings of
the lightest stops to the $h$ boson are proportional to $g_{h \tilde{t}_1
\tilde{t}_1} \sim A_t -\mu/\tb$, and large values of this couplings might have a
rather strong impact on the phenomenology of the MSSM Higgs bosons.\cite{rev} 
The measurement of this important coupling would open a window to probe
directly some of the soft--SUSY breaking terms of the potential. To measure
Higgs--stop couplings directly, one needs to consider the three--body
associated production of Higgs bosons with stop pairs [the SUSY analog to the
$t\bar{t}h$ associated production process].  At future linear $\ee$ colliders,
the final state $\tilde {t}_1 \tilde{t}_1 h$ may be generated in three
ways:\cite{jean-loic} $(i)$ two--body production of a mixed pair of top squarks
and the decay of the heaviest stop to the lightest one and a Higgs boson,
$(ii)$ the continuum production in $\ee$ annihilation $\ee \ra \tilde{t}_1
\tilde{t}_1h$ and $(iii)$ the continuum production in $\gamma \gamma$
collisions $\gamma \gamma \ra \tilde{t}_1 \tilde{t}_1h$. In the continuum
production in $\ee$ collisions at $\sqrt{s} \sim 800$ GeV, the cross sections
can exceed $1$ fb for not too large $\tilde{t}_1$ masses [$\lsim 200$ GeV] and
large values of the parameter $A_t$ [$\gsim 1$ TeV] and is thus comparable to
the one of SM--like process $\ee \ra t \bar{t}h$. This provides more than one
thousand events in a few years, with a luminosity  $\int {\cal L}\sim 500$
fb$^{-1}$, which should be sufficient to isolate the final state and measure
$g_{\tilde{t}_1 \tilde{t}_1 h}$  with some accuracy. Note that in most part of
the MSSM parameter space, the final state topology will consist of $4b$ quarks,
two of them peaking at an invariant mass $M_h$, two real or virtual $W$ bosons
and missing energy [i.e. the same topology as the process $\ee \ra t \bar{t}h$,
except for the missing energy].  

\vspace*{-2mm}

%\newpage

\section{Conclusions}

\vspace*{-1mm}

In the Standard Model, global fits of the electroweak data favor a light Higgs 
boson, $M_H \lsim 260$ GeV, and if the theory is to remain valid up to the GUT
scale, the Higgs boson should be lighter than $200$ GeV. In supersymmetric
extensions of the SM, there is always one light Higgs boson with a mass $M_h 
\lsim 135$ GeV in the minimal version and $M_h \lsim 205$ GeV in the most 
general one. Thus, a Higgs particle is definitely accessible at a linear
$\ee$ collider with a c.m. energy of $\sqrt{s} \gsim 350$ GeV. 

The detection of such a particle is not a problem at $\ee$ colliders. The
search can be made in a large variety of channels: Higgs--strahlung and vector
boson fusion processes in the SM, while additional processes are provided by
Higgs pair production in SUSY extensions. The cross sections give large samples
of events, especially if a very high luminosity, $\int {\cal L} \gsim 100$
fb$^{-1}$, is available. The signals are very clear in the clean environment of
$\ee$ colliders, and the possibility of making efficient $b$--tagging, using
missing mass techniques and the polarization of the initial beams, makes the
search even easier.  

The very high luminosities expected in some machines and the very clean
environment allow to investigate thoroughly the properties of the discovered
Higgs bosons. In the main production processes, the Higgs boson mass and width,
the spin and parity quantum numbers and the couplings to gauge bosons and
fermion can be measured. Higher order processes allow the direct determination
of some very important couplings such as the Higgs--$t\bar{t}$ Yukawa coupling,
the trilinear Higgs self--coupling, the couplings to photons and possibly, in
supersymmetric extensions of the SM,  the coupling to top squarks.

In conclusion: a future linear $\ee$ collider with a c.m. energy $\sqrt{s} 
\gsim 350$ GeV and a luminosity $\int {\cal L} \gsim 100$ fb$^{-1}$ is an
ideal instrument to search for Higgs bosons and to explore thoroughly the 
electroweak breaking mechanism. 

\vspace*{-2mm}

\section*{Acknowledgments}

\vspace*{-2mm}

I thank the members of the Higgs working group for discussions, and the
organizers for their invitation and for the nice and stimulating atmosphere 
of the meeting.

\end{document}